\documentclass[conference, letterpaper]{IEEEtran}

\usepackage{graphicx}

\usepackage[algo2e,linesnumbered,commentsnumbered, lined, boxed,ruled,vlined]{algorithm2e}

\SetAlFnt{\sf \scriptsize}
\SetKwSty{algokey}
\SetFuncSty{algofnt}
\SetArgSty{algofnt}
\SetCommentSty{algocomment}
\SetKwProg{Fn}{function}{}{}
\SetKwBlock{Begin}{ }{ }
\let\oldnl\nl
\newcommand{\nonl}{\renewcommand{\nl}{\let\nl\oldnl}}
\SetInd{1em}{1em}

\SetCommentSty{mycommfont}

\usepackage{xcolor}

\usepackage{listings}
\definecolor{cppColorBackground}{rgb}{0.99,0.99,0.99}
\definecolor{cppColorComment}{rgb}{0.04,0.29,0.04}
\definecolor{cppColorLine}{rgb}{0.6,0.6,0.6}
\definecolor{cppColorString}{rgb}{0,0.63,0}
\definecolor{cppColorKey}{rgb}{0.5,0.5,0}
\definecolor{cppColorDigit}{rgb}{0,0,0.5}
\usepackage{pgfplots}
\pgfplotsset{compat=newest}
\usetikzlibrary{patterns}

\usepackage{subcaption}	
\usepackage{color}

\definecolor{black6}{gray}{1.0}
\definecolor{black5}{gray}{0.82}
\definecolor{black4}{gray}{0.64}
\definecolor{black3}{gray}{0.46}
\definecolor{black2}{gray}{0.28}
\definecolor{black1}{gray}{0.10}

\definecolor{blue3}{rgb}{0.1,0.11,0.34}
\definecolor{blue2}{rgb}{0.18,0.34,0.46}
\definecolor{blue1}{rgb}{0.26,0.74,0.72}

\definecolor{green1}{rgb}{0.067,0.4,0.0067}
\definecolor{ForestGreen}{rgb}{0.0, 0.27, 0.13}

\definecolor{vecyellow}{rgb}{1.0, 0.8, 0.0}
\definecolor{vecgreen}{rgb}{0.1, 0.8, 0.0}

\usepackage{amsmath}
\usepackage{amssymb}

\usepackage{makecell}

\usepackage{placeins}

\pgfplotsset{
    /pgfplots/ybar legend/.style={
    /pgfplots/legend image code/.code={%
       \draw[##1,/tikz/.cd,yshift=-0.25em]
        (0cm,0cm) rectangle (3pt,0.8em);},
   },
}

\usepackage{fancyhdr}

\renewcommand{\thispagestyle}[2]{}

\fancypagestyle{plain}{
        \fancyhead{}
        \fancyhead[C]{first page center header}
        \fancyfoot{}
        \fancyfoot[C]{first page center footer}
}
\begin{document}

\title{A Novel Hybrid Quicksort Algorithm Vectorized using AVX-512 on Intel Skylake}
\author{\IEEEauthorblockN{Berenger Bramas}
\IEEEauthorblockA{Max Planck Computing and Data Facility (MPCDF) \\ Gießenbachstraße 2\\ 85748 Garching, Germany\\EMail: Berenger.Bramas@mpcdf.mpg.de}}

\newcommand{\todotext}[1]{\textbf{\textcolor{red}{TODO : #1}}}

\maketitle

\begin{abstract} The modern CPU's design, which is composed of hierarchical memory and SIMD/vectorization capability, governs the potential for algorithms to be transformed into efficient implementations.
The release of the AVX-512 changed things radically, and motivated us to search for an efficient sorting algorithm that can take advantage of it.
In this paper, we describe the best strategy we have found, which is a novel two parts hybrid sort, based on the well-known Quicksort algorithm. 
The central partitioning operation is performed by a new algorithm, and small partitions/arrays are sorted using a branch-free Bitonic-based sort.
This study is also an illustration of how classical algorithms can be adapted and enhanced by the AVX-512 extension.
We evaluate the performance of our approach on a modern Intel Xeon Skylake and assess the different layers of our implementation by sorting/partitioning integers, double floating-point numbers, and key/value pairs of integers. 
Our results demonstrate that our approach is faster than two libraries of reference: the GNU \emph{C++} sort algorithm by a speedup factor of 4, and the Intel IPP library by a speedup factor of 1.4.  


 \end{abstract}



\section{Introduction}



Sorting is a fundamental problem in computer science that always had the attention of the research community, because it is widely used to reduce the complexity of some algorithms.
Moreover, sorting is a central operation in specific applications such as, but not limited to, database servers~\cite{graefe2006implementing} and image rendering engines~\cite{bishop1998designing}. 
Therefore, having efficient sorting libraries on new architecture could potentially leverage the performance of a wide range of applications.  

The vectorization  --- that is, the CPU's capability to apply a single instruction on multiple data (SIMD) --- improves continuously, one CPU generation after the other.
While the difference between a scalar code and its vectorized equivalent was \textit{"only"} of a factor of 4 in the year 2000 (SSE), the difference is now up to a factor of 16 (AVX-512). 
Therefore, it is indispensable to \textit{vectorize} a code to achieve high-performance on modern CPUs, by using dedicated instructions and registers.
The conversion of a scalar code into a vectorized equivalent is straightforward for many classes of algorithms and computational kernels, and it can even be done with auto-vectorization for some of them.
However, the opportunity of vectorization is tied to the memory/data access patterns, such that data-processing algorithms (like sorting) usually require an important effort to be transformed.  
In addition, creating a fully vectorized implementation, without any scalar sections, is only possible and efficient if the instruction set provides the needed operations.
Consequently, new instruction sets, such as the AVX-512, allow for the use of approaches that were not feasible previously.

The Intel Xeon Skylake (SKL) processor is the second CPU that supports AVX-512, after the Intel Knight Landing. 
The SKL supports the AVX-512 instruction set~\cite{avx512ref}: it supports Intel AVX-512 foundational instructions (AVX-512F), Intel AVX-512 conflict detection instructions (AVX-512CD), Intel AVX-512 byte and word instructions (AVX-512BW), Intel AVX-512 doubleword and quadword instructions (AVX-512DQ), and Intel AVX-512 vector length extensions instructions (AVX-512VL). 
The AVX-512 not only allows work on SIMD-vectors of double the size, compared to the previous AVX(2) set, it also provides various new operations.

Therefore, in the current paper, we focus on the development of new sorting strategies and their efficient implementation for the Intel Skylake using AVX-512.
The contributions of this study are the following:   
\begin{itemize}
\item proposing a new partitioning algorithm using AVX-512,
\item defining a new Bitonic-sort variant for small arrays using AVX-512,
\item implementing a new Quicksort variant using AVX-512.
\end{itemize}
All in all, we show how we can obtain a fast and vectorized sorting algorithm
\footnote{
The functions described in the current study are available at https://gitlab.mpcdf.mpg.de/bbramas/avx-512-sort.
This repository includes a clean header-only library (branch master) and a test file that generates the performance study of the current manuscript (branch paper). 
The code is under MIT license.
}.

The rest of the paper is organized as follows:
Section~\ref{sec:brackground} gives background information related to vectorization and sorting.
We then describe our approach in Section~\ref{sec:sortingvec}, introducing our strategy for sorting small arrays, and the vectorized partitioning function, which are combined in our Quicksort variant.
Finally, we provide performance details in Section~\ref{sec:results} and the conclusion in Section~\ref{sec:conclusion}.

\section{Background}
\label{sec:brackground}

\subsection{Sorting Algorithms}

\subsubsection{Quicksort (QS) Overview}
\label{sec:qsintro}

\begin{figure}[h!]
\centering
\includegraphics[width=.7\columnwidth, keepaspectratio]{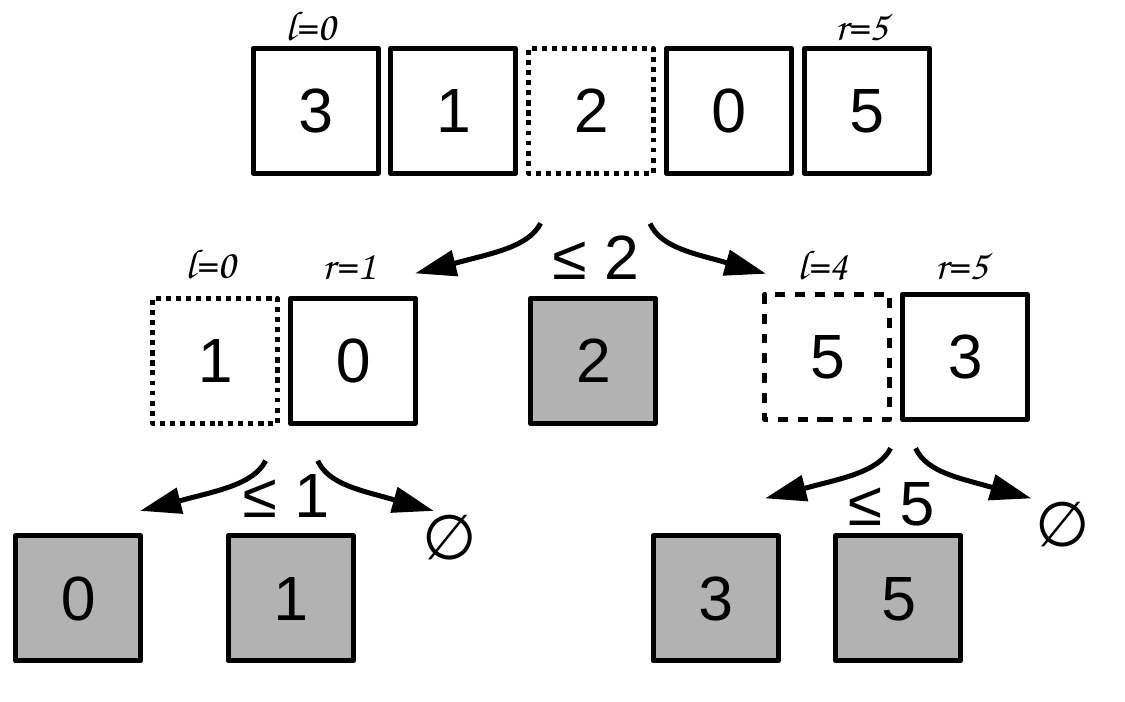}
\caption{Quicksort example to sort $[3,1,2,0,5]$ to $[0,1,2,3,5]$. The pivot is equal to the value in the middle: the first pivot is $2$, then at second recursion level it is $1$ and $5$.}
\label{fig:qsexample}
\end{figure}

QS was originally proposed in~\cite{hoare1962quicksort}.
It uses a \emph{divide-and-conquer} strategy, by recursively partitioning the input array, until it ends with partitions of one value.
The partitioning puts values lower than a \emph{pivot} at the beginning of the array, and greater values at the end, with a linear complexity.
QS has a worst-case complexity of O(\emph{n}$^2$), but an average complexity of O(\emph{n} log \emph{n}) in practice.
The complexity is tied to the choice of the partitioning pivot, which must be close to the median to ensure a low complexity.
However, its simplicity in terms of implementation, and its speed in practice, has turned it into a very popular sorting algorithm.
Figure~\ref{fig:qsexample} shows an example of a QS execution.

We provide in Appendix~\ref{algo:qs} the scalar QS algorithm.
Here, the term \textit{scalar} refers to a single value at the opposite of an SIMD vector.
In this implementation, the choice of the pivot is naively made by selecting the value in the middle before partitioning, and this can result in very unbalanced partitions.
This is why more advanced heuristics have been proposed in the past, like selecting the median from several values, for example.

\subsubsection{GNU std::sort Implementation (STL)}

The worst case complexity of QS makes it no longer suitable to be used as a standard \emph{C++} sort.
In fact, a complexity of O(\emph{n} log \emph{n}) in average was required until year 2003~\cite{cpp03}, but it is now a worst case limit~\cite{cpp14} that a pure QS implementation cannot guarantee.
Consequently, the current implementation is a 3-part hybrid sorting algorithm \textit{i}.\textit{e}.\ it relies on 3 different algorithms~\footnote{See the libstdc++ documentation on the sorting algorithm available at https://gcc.gnu.org/onlinedocs/libstdc++/libstdc++-html-USERS-4.4/a01347.html\#l05207}.
The algorithm uses an Introsort~\cite{musser1997introspective} to a maximum depth of 2 $\times$ log$^2$ n to obtain small partitions that are then sorted using an insertion sort.
Introsort is itself a 2-part hybrid of Quicksort and heap sort.

\subsubsection{Bitonic Sorting Network}

\begin{figure}
\centering
\begin{subfigure}[c]{0.54\columnwidth}
\includegraphics[width=\textwidth, keepaspectratio]{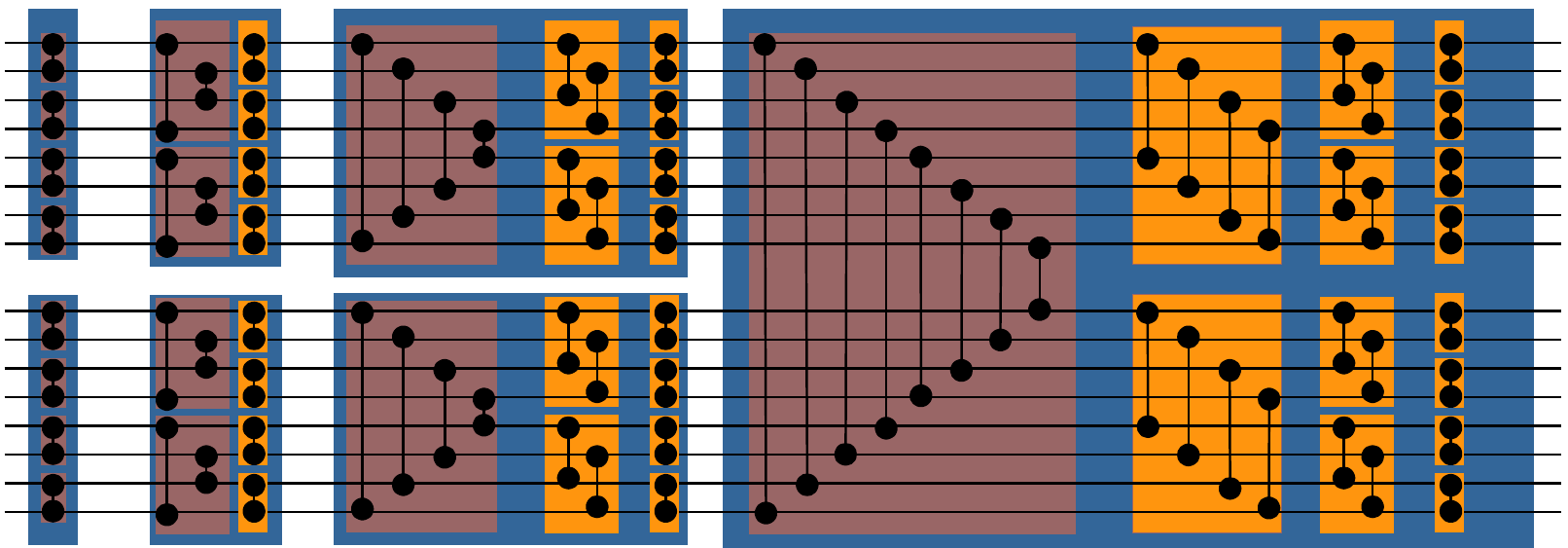}
\caption{Bitonic sorting network for input of size 16. All vertical bars/switches exchange values in the same direction.}
\label{fig:bitonicnetwork}
\end{subfigure}
\begin{subfigure}[c]{0.4\columnwidth}
\includegraphics[width=\textwidth, keepaspectratio]{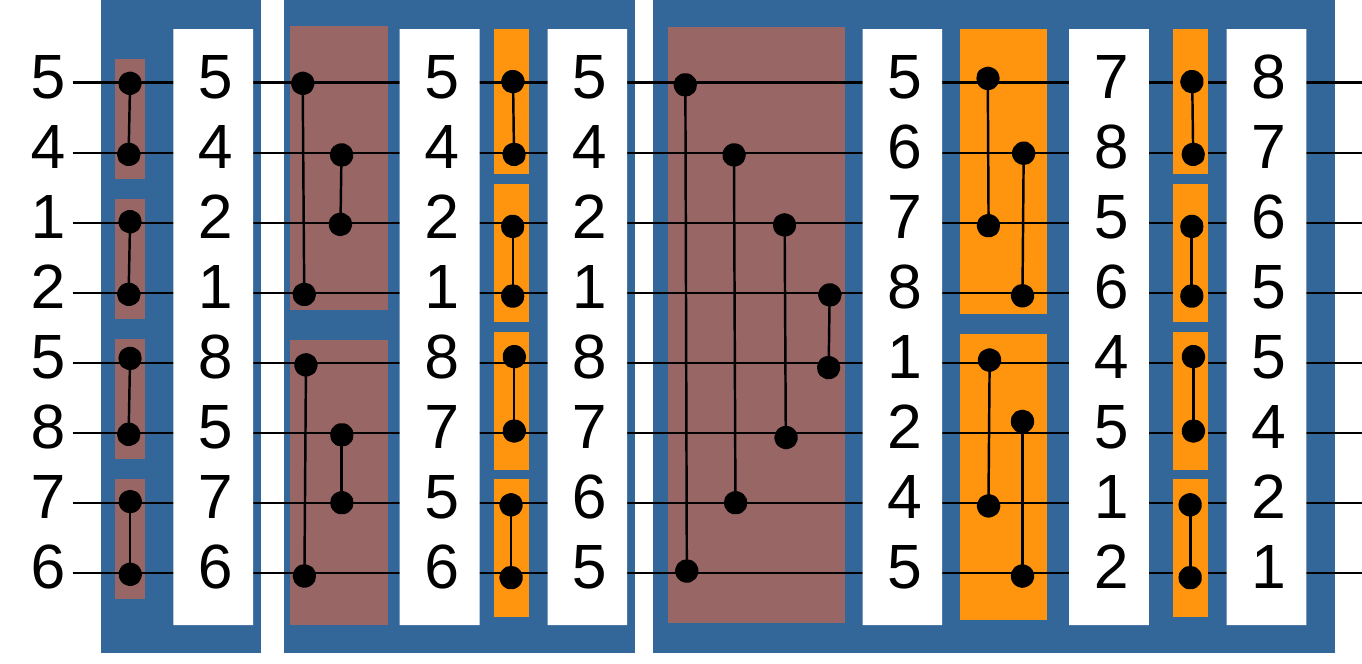}
\caption{Example of 8 values sorted by a Bitonic sorting network.\\}
\label{fig:bitonicex}
\end{subfigure}
\caption{Bitonic sorting network examples. In red boxes, the exchanges are done from extremities to the center. Whereas in orange boxes, the exchanges are done with a linear progression.}
\label{fig:bitonicall}
\end{figure}

In computer science, a sorting network is an abstract description of how to sort a fixed number of values \textit{i}.\textit{e}.\ how the values are compared and exchanged.
This can be represented graphically, by having each input value as a horizontal line, and each \textit{compare and exchange} unit as a vertical connection between those lines.
There are various examples of sorting networks in the literature, but we concentrate our description on the Bitonic sort from~\cite{batcher1968sorting}.
This network is easy to implement and has an algorithm complexity of O(\emph{n} log(\emph{n})$^2$ ).
It has demonstrated good performances on parallel computers~\cite{nassimi1979bitonic} and GPUs~\cite{owens2008gpu}.
Figure~\ref{fig:bitonicnetwork} shows a Bitonic sorting network to process 16 values.
A sorting network can be seen as a time line, where input values are transferred from left to right, and exchanged if needed at each vertical bar.
We illustrate an execution in Figure~\ref{fig:bitonicex}, where we print the intermediate steps while sorting an array of 8 values.
The Bitonic sort is not stable because it does not maintain the original order of the values.

If the size of the array to sort is known, it is possible to implement a sorting network by hard-coding the connections between the lines.
This can be seen as a direct mapping of the picture.
However, when the array size is unknown, the implementation can be made more flexible by using a formula/rule to decide when to compare/exchange values.

\subsection{Vectorization}

\begin{figure}
\centering
\includegraphics[width=.7\columnwidth, keepaspectratio]{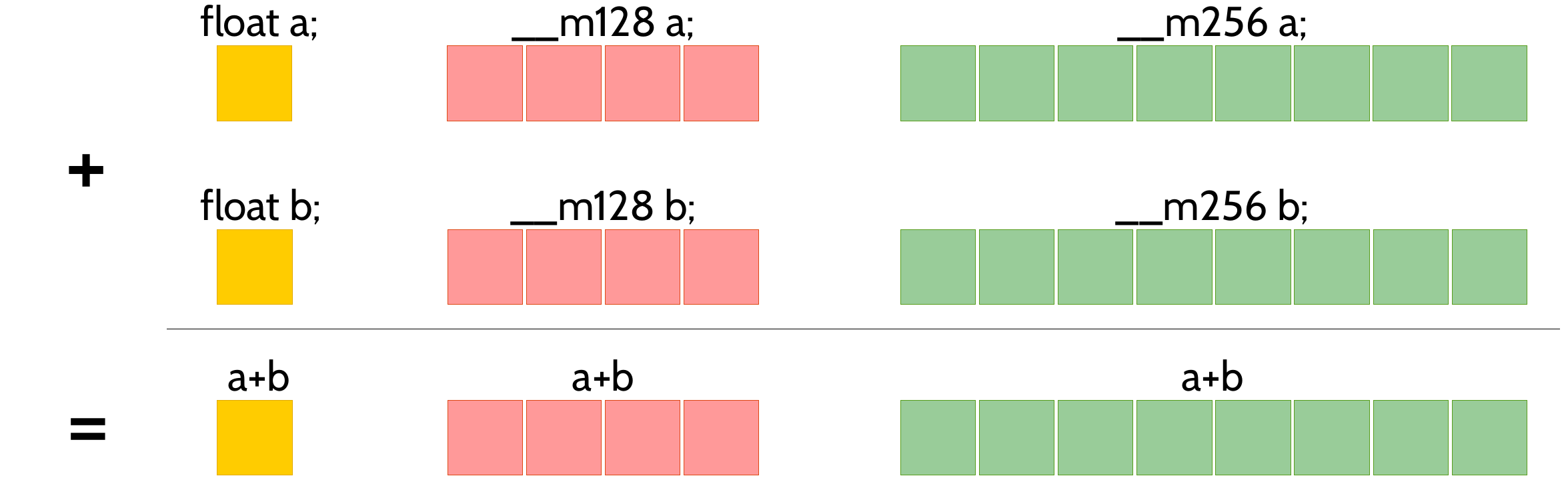}
\caption{Summation example of single precision floating-point values using : ($\color{vecyellow} \blacksquare$) scalar standard \textit{C++} code, ($\color{red} \blacksquare$) SSE SIMD-vector of $4$ values , ($\color{vecgreen} \blacksquare$) AVX SIMD-vector of $8$ values.}
\label{fig:vec}
\end{figure}

The term vectorization refers to a CPU's feature to apply a single operation/instruction to a vector of values instead of only a single value~\cite{kogge1981architecture}.
It is common to refer to this concept by Flynn's taxonomy term, SIMD, for single instruction on multiple data.
By adding SIMD instructions/registers to CPUs, it has been possible to increase the peak performance of single cores, despite the stagnation of the clock frequency. 
The same strategy is used on new hardware, where the length of the SIMD registers has continued to increase.
In the rest of the paper, we use the term \emph{vector} for the data type managed by the CPU in this sense.
It has no relation to an expandable vector data structure, such as \emph{std::vector}.
The size of the vectors is variable and depends on both the instruction set and the type of vector element, and corresponds to the size of the registers in the chip.
Vector extensions to the \emph{x86} instruction set, for example, are SSE~\cite{sseref}, AVX~\cite{avxref}, and AVX512~\cite{avx512ref}, which support vectors of size 128, 256 and 512 bits respectively.
This means that an SSE vector is able to store four single precision floating point numbers or two double precision values.
Figure~\ref{fig:vec} illustrates the difference between a scalar summation and a vector summation for SSE or AVX, respectively.
An AVX-512 SIMD-vector is able to store 8 double precision floating-point numbers or 16 integer values, for example.
Throughout this document, we use \textit{intrinsic} function extension instead of the assembly language to write vectorized code on top of the AVX-512 instruction set.
Intrinsics are small functions that are intended to be replaced with a single assembly instruction by the compiler.

\subsubsection{AVX-512 Instruction Set}
\label{sec:avx512instructions}

As previous \emph{x86} vectorization extensions, the AVX-512 has instructions to load a contiguous block of values from the main memory and to transform it into a SIMD-vector (load).
It is also possible to fill a SIMD-vector with a given value (set), and move back a SIMD-vector into memory (store).
A permutation instruction allows to re-order the values inside a SIMD-vector using a second integer array which contains the permutation indexes.
This operation was possible in since AVX/AVX2 using \emph{permutevar8x32} (instruction vperm(d,ps)).
The instructions \emph{vminpd/vpminsd} return a SIMD-vector where each value correspond to the minimum of the values from the two input vectors at the same position.
It is possible to obtain the maximum with instructions \emph{vpmaxsd/vmaxpd}.

In AVX-512, the value returned by a test/comparison (\emph{vpcmpd/vcmppd}) is a mask (integer) and not an SIMD-vector of integers, as it was in SSE/AVX.
Therefore, it is easy to modify and work directly on the mask with arithmetic and binary operations for scalar integers.
Among the mask-based instructions, the \emph{mask move (vmovdqa32/vmovapd)} allows for the selection of values between two vectors, using a mask.
Achieving the same result was possible in previous instruction sets using the \emph{blend} instruction since SSE4, and using several operations with previous instruction sets.

The AVX-512 provides operations that do not have an equivalent in previous extensions of the \emph{x86} instruction sets, such as the \emph{store-some} (\emph{vpcompressps/vcompresspd}) and \emph{load-some} (\emph{vmovups/vmovupd}).
The \emph{store-some} operation allows to save only a part of a SIMD-vector into memory.
Similarly, the \emph{load-some} allows to load less values than the size of a SIMD-vector from the memory.
The values are loaded/saved contiguously.
This is a major improvement, because without this instruction, several operations are needed to obtain the same result.
For example, to save some values from a SIMD-vector \emph{v} at address \emph{p} in memory, one possibility is to load the current values from \emph{p} into a SIMD-vector \emph{v'}, permute the values in \emph{v} to move the values to store at the beginning, merge \emph{v} and \emph{v'}, and finally save the resulting vector.

\subsection{Related Work on Vectorized Sorting Algorithms}

The literature on sorting and vectorized sorting implementations is extremely large.
Therefore, we only cite some of the studies that we consider most related to our work.

The sorting technique from~\cite{sanders2004super} tries to remove branches and improves the prediction of a scalar sort, and they show a speedup by a factor of 2 against the STL (the implementation of the STL was different at that time). 
This study illustrates the early strategy to adapt sorting algorithms to a given hardware, and  also shows the need for low-level optimizations, due to the limited instructions available at that time.

In~\cite{inoue2007aa}, the authors propose a parallel sorting on top of combosort vectorized with the VMX instruction set of IBM architecture.
Unaligned memory access is avoided, and the L2 cache is efficiently managed by using an out-of-core/blocking scheme.
The authors show a speedup by a factor of 3 against the GNU \emph{C++} STL. 

In~\cite{furtak2007using}, the authors use a sorting-network for small-sized arrays, similar to our own approach. 
However, instead of dividing the main array into sorted partitions (partitions of increasing contents), and applying a small efficient sort on each of those partitions,
the authors perform the opposite.
They apply multiple small sorts on sub-parts of the array, and then they finish with a complicated merge scheme using extra memory to globally sort all the sub-parts.
A very similar approach was later proposed in~\cite{chhugani2008efficient}. 

The recent work in~\cite{gueron2016fast} targets AVX2. 
The authors use a Quicksort variant with a vectorized partitioning function, and an insertion sort once the partitions are small enough (as the STL does).
The partition method relies on look-up tables, with a mapping between the comparison's result of an SIMD-vector against the pivot, and the move/permutation that must be applied to the vector.
The authors demonstrate a speedup by a factor of 4 against the STL, but their approach is not always faster than the Intel IPP library.
The proposed method is not suitable for AVX-512 because the lookup tables will occupy too much memory.
This issue, as well as the use of extra memory, can be solved with the new instructions of the AVX-512.
As a side remark, the authors do not compare their proposal to the standard \emph{C++ partition} function, even so, it is the only part of their algorithm that is vectorized.

\section{Sorting with AVX-512}
\label{sec:sortingvec}


\subsection{Bitonic-Based Sort on AVX-512 SIMD-Vectors}
\label{sec:bitonicsort}

In this section, we describe our method to sort small arrays that contain less than 16 times \emph{VEC\_SIZE}, where \emph{VEC\_SIZE} is the number of values in a SIMD-vector.
This function is later used in our final QS implementation to sort small enough partitions.

\subsubsection{Sorting one SIMD-vector}

To sort a single vector, we perform the same operations as the ones shown in Figure~\ref{fig:bitonicnetwork}: we compare and exchange values following the indexes from the Bitonic sorting network.
However, thanks to the vectorization, we are able to work on the entire vector without having to iterate on the values individually.
We know the positions that we have to compare and exchange at the different stages of the algorithm.
This is why, in our approach, we rely on static (hard-coded) permutation vectors, as shown in Algorithm~\ref{algo:bitonic1v}.
In this algorithm, the \emph{compare\_and\_exchange} function performs all the \emph{compare and exchange} that are applied at the same time in the Bitonic algorithm \textit{i}.\textit{e}. the operations that are at the same horizontal position in the figure.
To have a fully vectorized function, we implement the \emph{compare\_and\_exchange} in three steps.
First, we permute the input vector \emph{v} into \emph{v'} with the given permutation indexes \emph{p}.
Second, we obtain two vectors \emph{w$_{min}$} and \emph{w$_{max}$} that contain the minimum and maximum values between both \emph{v} and \emph{v'}.
Finally, we selects the values from \emph{w$_{min}$} and \emph{w$_{max}$} with a mask-based move, where the mask indicates in which direction the exchanges have to be done.
The \emph{C++} source code of a fully vectorized branch-free implementation is given in Appendix~\ref{app:code} (Code~\ref{code:bitonic1v}).

\begin{algorithm2e}[h!]
\SetAlgoLined
\DontPrintSemicolon
\KwIn{vec: a double floating-point AVX-512 vector to sort.}
\KwOut{vec: the vector sorted.}
 \Fn{simd\_bitonic\_sort\_1v(vec)}
 {
        compare\_and\_exchange(vec, [6, 7, 4, 5, 2, 3, 0, 1])\;
        compare\_and\_exchange(vec, [4, 5, 6, 7, 0, 1, 2, 3])\;
        compare\_and\_exchange(vec, [6, 7, 4, 5, 2, 3, 0, 1])\;
        compare\_and\_exchange(vec, [0, 1, 2, 3, 4, 5, 6, 7])\;
        compare\_and\_exchange(vec, [5, 4, 7, 6, 1, 0, 3, 2])\;
        compare\_and\_exchange(vec, [6, 7, 4, 5, 2, 3, 0, 1])\;
 }
 \caption{SIMD Bitonic sort for one vector of double floating-point values.}
 \label{algo:bitonic1v}
\end{algorithm2e}

\subsubsection{Sorting more than one SIMD-vectors}
\label{sec:multiplebitonicsort}

The principle of using static permutation vectors to sort a single SIMD-vector can be applied to sort several SIMD-vectors.
In addition, we can take advantage of the repetitive pattern of the Bitonic sorting network to re-use existing functions.
More precisely, to sort $V$ vectors, we re-use the function to sort $V/2$ vectors and so on.
We provide an example to sort two SIMD-vectors in Algorithm~\ref{algo:bitonic2v}, where we start by sorting each SIMD-vector individually using the \emph{bitonic\_simd\_sort\_1v} function.
Then, we compare and exchange values between both vectors (line 5), and finally applied the same operations on each vector individually (lines 6 to 11).
In our sorting implementation, we provide the functions to sort up to 16 SIMD-vectors, which correspond to 256 integer values or 128 double floating-point values.

\begin{algorithm2e}[h!]
\SetAlgoLined
\DontPrintSemicolon
\KwIn{vec1 and vec2: two double floating-point AVX-512 vectors to sort.}
\KwOut{vec1 and vec2: the two vectors sorted with vec1 lower or equal than vec2.}
 \Fn{simd\_bitonic\_sort\_2v(vec1, vec2)}
 {
        \tcp{Sort each vector using bitonic\_simd\_sort\_1v}
        simd\_bitonic\_sort\_1v(vec1)\;
        simd\_bitonic\_sort\_1v(vec2)\;

        compare\_and\_exchange\_2v(vec1, vec2, [0, 1, 2, 3, 4, 5, 6, 7])\;
        compare\_and\_exchange(vec1, [3, 2, 1, 0, 7, 6, 5, 4])\;
        compare\_and\_exchange(vec2, [3, 2, 1, 0, 7, 6, 5, 4])\;
        compare\_and\_exchange(vec1, [5, 4, 7, 6, 1, 0, 3, 2])\;
        compare\_and\_exchange(vec2, [5, 4, 7, 6, 1, 0, 3, 2])\;
        compare\_and\_exchange(vec1, [6, 7, 4, 5, 2, 3, 0, 1])\;
        compare\_and\_exchange(vec2, [6, 7, 4, 5, 2, 3, 0, 1])\;
 }
 \caption{SIMD bitonic sort for two vectors of double floating-point values.}
 \label{algo:bitonic2v}
\end{algorithm2e}


\subsubsection{Sorting Small Arrays}
\label{sec:anysize}

Each of our simd-Bitonic-sort functions are designed for a specific number of SIMD-vectors.
However, we intend to sort arrays that do not have a size multiple of the SIMD-vector's length, because they are obtained from the partitioning stage of the QS.
Consequently, when we have to sort a small array, we first load it into SIMD-vectors, and then, we pad the last vector with the greatest possible value.
This guarantee that the padding values have no impact on the sorting results by staying at the end of the last vector.
The selection of appropriate simd-Bitonic-sort function, that matches the size of the array to sort, can be done efficiently with a switch statement.
In the following, we refer to this interface as the \emph{simd\_bitonic\_sort\_wrapper} function.

\subsection{Partitioning with AVX-512}
\label{sec:partition512}

Algorithm~\ref{alg:simdpartition} shows our strategy to develop a vectorized partitioning method.
This algorithm is similar to a scalar partitioning function: there are iterators that start from both extremities of the array to keep track of where to load/store the values, and the process stops when some of these iterators meet.
In its steady state, the algorithm loads an SIMD-vector using the left or right indexes (at lines 19 and 24), and partitions it using the \emph{partition\_vec} function (at line 27).
The \emph{partition\_vec} function compares the input vector to the pivot vector (at line 47), and stores the values --- lower or greater --- directly in the array using a store-some instruction (at lines 51 and 55).
The store-some is an AVX-512 instruction that we described in Section~\ref{sec:avx512instructions}.
The initialization of our algorithm starts by loading one vector from each array's extremities to ensure that no values will be overwritten during the steady state (lines 12 and 16).
This way, our implementation works in-place and only needs three SIMD-vectors.
Algorithm~\ref{alg:simdpartition} also includes, as side comments, possible optimizations in case the array is more likely to be already partitioned (A), or to reduce the data displacement of the values (B).
The AVX-512 implementation of this algorithm is given in Appendix~\ref{app:code} (Code~\ref{code:simdpartition512}).
One should note that  we use a scalar partition function if there are less than 2 $\times$ VEC\_SIZE values in the given array (line 3).


\begin{algorithm2e}[h!]    
\SetAlgoLined
\DontPrintSemicolon
\KwIn{array: an array to partition. length: the size of array. pivot: the reference value}
\KwOut{array: the array partitioned. left\_w: the index between the values lower and larger than the pivot.}      
\Fn{simd\_partition(array, length, pivot)}
{
   \tcp{If too small use scalar partitioning}
   \If{length $\le 2 \times VEC\_SIZE$}
   {
        Scalar\_partition(array, length)\;
        return \;
   }

   \tcp{Set: Fill a vector with all values equal to pivot}
   pivotvec = simd\_set\_from\_one(pivot)\;

   \tcp{Init iterators and save one vector on each extremity}
   left = 0\;
   left\_w = 0\;
   left\_vec = simd\_load(array, left)\;
   left = left + VEC\_SIZE\;

   right = length-VEC\_SIZE\;
   right\_w = length\;
   right\_vec = simd\_load(array, right)\;

   \While{left + VEC\_SIZE $\le$ right}
   {
        \eIf{ (left - left\_w) $\le$ (right\_w - right) }
        {
            val = simd\_load(array, left)\;
            left = left + VEC\_SIZE\;
            \tcp{(B) Possible optimization, swap val and left\_vec}
        }
        {
            right = right - VEC\_SIZE\;
            val = simd\_load(array, right)\;
            \tcp{(B) Possible optimization, swap val and right\_vec}
        }

        [left\_w, right\_w] = partition\_vec(array, val, pivotvec, left\_w, right\_w)\;
    }
    \tcp{Process left\_val and right\_val}
    [left\_w, right\_w] = partition\_vec(array, left\_val, pivotvec, left\_w, right\_w)\;
    [left\_w, right\_w] = partition\_vec(array, right\_val, pivotvec, left\_w, right\_w)\;

    \tcp{Proceed remaining values (less than VEC\_SIZE values)}
    nb\_remaining = right - left\;
    val = simd\_load(array, left)\;
    left = right\;

    mask = get\_mask\_less\_equal(val, pivotvec)\;

    mask\_low = cut\_mask(mask, nb\_remaining)\;
    mask\_high = cut\_mask(reverse\_mask(mask) , nb\_remaining)\;

    \tcp{(A) Possible optimization, do only if mask\_low is not 0}
    simd\_store\_some(array, left\_w, mask\_low, val)\;
    left\_w = left\_w + mask\_nb\_true(mask\_low)\;

    \tcp{(A) Possible optimization, do only if mask\_high is not 0}
    right\_w = right\_w - mask\_nb\_true(mask\_high)\;
    simd\_store\_some(array, right\_w, mask\_high, val)\;

    return left\_w\;
}
\Fn{partition\_vec(array, val, pivotvec, left\_w, right\_w)}
{
    mask = get\_mask\_less\_equal(val, pivotvec)\;

    nb\_low = mask\_nb\_true(mask)\;
    nb\_high = VEC\_SIZE-nb\_low\;

    \tcp{(A) Possible optimization, do only if mask is not 0}
    simd\_store\_some(array, left\_w, mask, val)\;
    left\_w = left\_w + nb\_low\;

    \tcp{(A) Possible optimization, do only if mask is not all true}
    right\_w = right\_w - nb\_high\;
    simd\_store\_some(array, right\_w, reverse\_mask(mask), val)\;

    return [left\_w, right\_w]\;
}

\caption{SIMD partitioning. $VEC\_SIZE$ is the number of values inside a SIMD-vector of type array's elements.}
\label{alg:simdpartition}
\end{algorithm2e}

\subsection{Quicksort Variant}

Our QS is given in Algorithm~\ref{algo:simdqs}, where we partition the data using the \emph{simd\_partition} function from Section~\ref{sec:partition512}, and then sort the small partitions using the \emph{simd\_bitonic} \emph{\_sort\_wrapper} function from Section~\ref{sec:bitonicsort}.
The obtained algorithm is very similar to the scalar QS given in Appendix~\ref{algo:qs}.

\begin{algorithm2e}[h!]
\SetAlgoLined
\DontPrintSemicolon
\KwIn{array: an array to sort. length: the size of array.}
\KwOut{array: the array sorted.}
 \Fn{simd\_QS(array, length)}
 {
    simd\_QS\_core(array, 0, length-1)\;
 }
\BlankLine
 \Fn{simd\_QS\_core(array, left, right)}
 {
    \tcp{Test if we must partition again or if we can sort}
    \eIf{ left $+$ SORT\_BOUND $<$ right }
    {
        pivot\_idx = select\_pivot\_pos(array, left, right)\;
        swap(array[pivot\_idx], array[right])\;
        partition\_bound = simd\_partition(array, left, right, array[right])\;
        swap(array[partition\_bound], array[right])\;
        simd\_QS\_core(array, left, partition\_bound-1)\;
        simd\_QS\_core(array, partition\_bound+1, right)\;
    }
    {
        simd\_bitonic\_sort\_wrapper(sub\_array(array, left), right-left+1)\;
    }
 }
 \caption{SIMD Quicksort. $select\_pivot\_pos$ returns a pivot.}
 \label{algo:simdqs}
\end{algorithm2e}

\subsection{Sorting Key/Value Pairs}

The previous sorting methods are designed to sort an array of numbers.
However, some applications need to sort key/value pairs.
More precisely, the sort is applied on the keys, and the values contain extra information and could be pointers to arbitrary data structures, for example.
Storing each key/value pair contiguously in memory is not adequate for vectorization because it requires transforming the data.
Therefore, in our approach, we store the keys and the values in two distinct arrays.
To extend the simd-Bitonic-sort and simd-partition functions, we must ensure that the same permutations/moves are applied to the keys and the values.
For the partition function, this is trivial.
The same mask is used in combination with the store-some instruction for both arrays.
For the Bitonic-based sort, we manually apply the permutations that were done on the vector of keys to the vector of values.
To do so, we first save the vector of keys \emph{k} before it is permuted by a \emph{compare and exchange}, using the Bitonic permutation vector of indexes \emph{p}, into \emph{k'}.
We compare \emph{k} and \emph{k'} to obtain a mask \emph{m} that expresses what moves have been done.
Then, we permute our vector of values \emph{v} using \emph{p} into \emph{v'}, and we select the correct values between \emph{v} and \emph{v'} using \emph{m}.
Consequently, we perform this operation at the end of the \emph{compare\_and\_exchange} in all the Bitonic-based sorts.

\section{Performance Study}
\label{sec:results}

\subsection{Configuration}

We asses our method on an Intel(R) Xeon(R) Platinum 8170 \emph{Skylake} CPU at 2.10GHz, with caches of sizes 32K-Bytes, 1024K-Bytes and 36608K-Bytes, at levels L1, L2 and L3 respectively.
The process and allocations are bound with \textit{numactl --physcpubind=0 --localalloc}.
We use the Intel compiler 17.0.2 (20170213) with aggressive optimization flag \emph{-O3}.

We compare our implementation against two references.
The first one is the GNU STL 3.4.21 from which we use the \emph{std::sort} and \emph{std::partition} functions.
The second one is the Intel Integrated Performance Primitives (IPP) 2017 which is a library optimized for Intel processors.
We use the IPP radix-based sort (function \emph{ippsSortRadixAscend\_[type]\_I}).
This function require additional space, but it is known as one of the fastest existing sorting implementation.

The test file used for the following benchmark is available online\footnote{
The test file that generates the performance study is available at https://gitlab.mpcdf.mpg.de/bbramas/avx-512-sort (branch paper) under MIT license.
}, it includes the different sorts presented in this study plus some additional strategies and tests.
Our simd-QS uses a 3-values median pivot selection (similar to the STL sort function).
The arrays to sort are populated with randomly generated values.

\subsection{Performance to Sort Small Arrays}

\begin{figure}
\begin{subfigure}[c]{\columnwidth}
        \begin{tikzpicture}
        \begin{axis}[
            ymode=log,log basis y=10,
            height=.2\textheight,
            width=.97\textwidth,
            xmin = 0,
            xmax = 256,
            ymax=1.5E-7,
            ymin=1E-10,
            axis lines*=left,
            legend style={at={(1,1.2)}, anchor=east,legend columns=3, font=\scriptsize},
            ylabel={Time in s/ n log(n)},
            xlabel={Number of values n},
            yticklabel style = {font=\scriptsize,xshift=0.5ex},
            xticklabel style = {font=\scriptsize,yshift=0.5ex},
            ylabel style = {font=\scriptsize},
            xlabel style = {font=\scriptsize},
            every node near coord/.append style={font=\scriptsize, color=gray,yshift=+1pt, /pgf/number format/.cd,fixed zerofill,precision=0},
                 every axis legend/.append style={nodes={right}},
            ]
\addplot[draw=purple] table [x index=0, y index=2, header=true] {./resultsres-intel1smallres-int.data}; 
\addplot[draw=brown] table [x index=0, y index=6, header=true] {./resultsres-intel1smallres-int.data}; 
\addplot[draw=ForestGreen] table [x index=0, y index=4, header=true] {./resultsres-intel1smallres-int.data}; 
\addplot[draw=ForestGreen, only marks, mark=o,nodes near coords,point meta=explicit symbolic] table [x index=0, y index=4, meta index=7, header=true] {./resultsres-intel1smallres-int-speedup.data};

\legend{std::sort, Intel IPP, AVX-512-bitonic sort}

        \end{axis}
        \end{tikzpicture}
    \caption{Integer (int)}
    \label{fig:smallint}
    \end{subfigure}
    \hfill
    \begin{subfigure}[c]{\columnwidth}
    \begin{tikzpicture}
    \begin{axis}[
            ymode=log,log basis y=10,
            height=.2\textheight,
            width=.97\textwidth,
            xmin = 0,
            xmax = 128,
            ymax=1.5E-7,
            ymin=1E-10,
            axis lines*=left,
            legend style={at={(1.7,0.5)}, anchor=east,legend columns=1, font=\scriptsize},
            ylabel={Time in s/ n log(n)},
            xlabel={Number of values n},
            yticklabel style = {font=\scriptsize,xshift=0.5ex},
            xticklabel style = {font=\scriptsize,yshift=0.5ex},
            ylabel style = {font=\scriptsize},
            xlabel style = {font=\scriptsize},
            every node near coord/.append style={font=\scriptsize, color=gray,yshift=+1pt, /pgf/number format/.cd,fixed zerofill,precision=0},
                 every axis legend/.append style={nodes={right}},
        ]

\addplot[draw=purple] table [x index=0, y index=2, header=true] {./resultsres-intel1smallres-double.data}; 
\addplot[draw=brown] table [x index=0, y index=6, header=true] {./resultsres-intel1smallres-double.data}; 
\addplot[draw=ForestGreen] table [x index=0, y index=4, header=true] {./resultsres-intel1smallres-double.data}; 
\addplot[draw=ForestGreen, only marks, mark=o,nodes near coords,point meta=explicit symbolic] table [x index=0, y index=4, meta index=7, header=true] {./resultsres-intel1smallres-double-speedup.data};

    \end{axis}
    \end{tikzpicture}
        \caption{Floating-point (double)}
    \end{subfigure}
    \hfill
    \begin{subfigure}[c]{\columnwidth}
    \begin{tikzpicture}
    \begin{axis}[
            ymode=log,log basis y=10,
            height=.2\textheight,
            width=.97\textwidth,
            xmin = 0,
            xmax = 256,
            ymax=1.5E-7,
            ymin=1E-10,
            axis lines*=left,
            legend style={at={(1.7,0.5)}, anchor=east,legend columns=1, font=\scriptsize},
            ylabel={Time in s/ n log(n)},
            xlabel={Number of values n},
            yticklabel style = {font=\scriptsize,xshift=0.5ex},
            xticklabel style = {font=\scriptsize,yshift=0.5ex},
            ylabel style = {font=\scriptsize},
            xlabel style = {font=\scriptsize},
            every node near coord/.append style={font=\scriptsize, color=gray,yshift=+1pt, /pgf/number format/.cd,fixed zerofill,precision=0},
                 every axis legend/.append style={nodes={right}},
        ]

\addplot[draw=purple] table [x index=0, y index=2, header=true] {./resultsres-intel1smallres-pair-int.data}; 
\addplot[draw=ForestGreen] table [x index=0, y index=4, header=true] {./resultsres-intel1smallres-pair-int.data}; 
\addplot[draw=ForestGreen, only marks, mark=o,nodes near coords,point meta=explicit symbolic] table [x index=0, y index=4, meta index=5, header=true] {./resultsres-intel1smallres-pair-int-speedup.data};

    \end{axis}
    \end{tikzpicture}
    \caption{Key/value integer pair (int[2])}
    \label{fig:smallpair}
    \end{subfigure}
    \caption{Execution time divided by \emph{n log(n)} to sort from 1 to 16 $\times$ \emph{VEC\_SIZE} values.
            The execution time is obtained from the average of 10$^4$ sorts for each size.
            The speedup of the AVX-512-bitonic against the fastest between STL and IPP is shown above the AVX-512-bitonic line.}
\label{res:variablesize}
\end{figure}
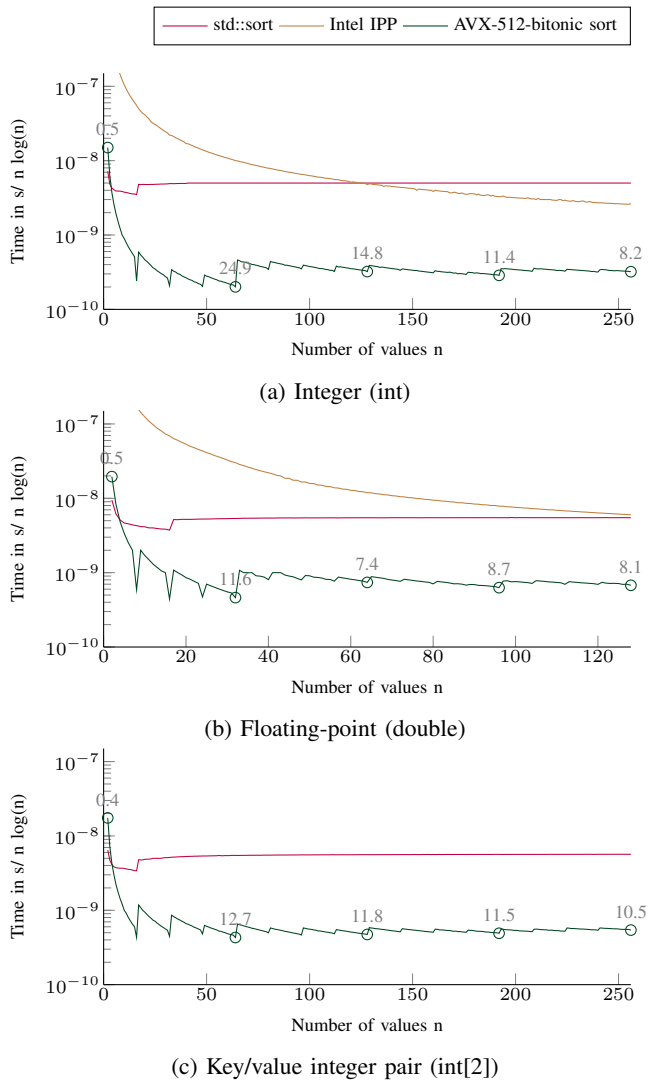

Figure~\ref{res:variablesize} shows the execution times to sort arrays of size from 1 to 16 $\times$ \emph{VEC\_SIZE}, which corresponds to 128 double floating-point values, or 256 integer values.
We also test arrays of size not multiple of the SIMD-vector's length.
The AVX-512-bitonic always delivers better performance than the Intel IPP for any size, and better performance than the STL when sorting more than 5 values.
The speedup is significant, and is around 8 in average.
The execution time per item increases every \emph{VEC\_SIZE} values because the cost of sorting is not tied to the number of values but to the number of SIMD-vectors to sort, as explained in Section~\ref{sec:anysize}.
For example, in Figure~\ref{fig:smallint}, the execution time to sort 31 or 32 values is the same, because we sort one SIMD-vector of 32 values in both cases.
Our method to sort key/value pairs seems efficient, see Figure~\ref{fig:smallpair}, because the speedup is even better against the STL compared to the sorting of integers.

\subsection{Partitioning Performance}

\begin{figure}
\begin{subfigure}[c]{\columnwidth}
        \begin{tikzpicture}
        \begin{axis}[
            xmode=log,log basis x=10,
            ymode=log,log basis y=10,
            height=.2\textheight,
            width=.97\textwidth,
            xmin=32,
            xmax=1073741824,
            ymax=1.5E-8,
            ymin=1E-10,
            axis lines*=left,
            legend style={at={(1,1.2)}, anchor=east,legend columns=3, font=\scriptsize},
            ylabel={Time in s/ n },
            xlabel={Number of values n},
            yticklabel style = {font=\scriptsize,xshift=0.5ex},
            xticklabel style = {font=\scriptsize,yshift=0.5ex},
            ylabel style = {font=\scriptsize},
            xlabel style = {font=\scriptsize},
            every node near coord/.append style={font=\scriptsize, color=gray,yshift=+1pt, /pgf/number format/.cd,fixed zerofill,precision=0},
                 every axis legend/.append style={nodes={right}},
            ]
\addplot[draw=purple] table [x index=0, y index=2, header=true] {./resultsres-intel1partitions-int.data}; 
\addplot[draw=ForestGreen] table [x index=0, y index=6, header=true] {./resultsres-intel1partitions-int.data}; 
\addplot[draw=ForestGreen, only marks, mark=o,nodes near coords,point meta=explicit symbolic] table [x index=0, y index=6, meta index=7, header=true] {./resultsres-intel1partitions-int.data};

\legend{std::partition, AVX-512-partition}

        \end{axis}
        \end{tikzpicture}
    \caption{Integer (int)}
    \end{subfigure}
    \hfill
    \begin{subfigure}[c]{\columnwidth}
    \begin{tikzpicture}
    \begin{axis}[
            xmode=log,log basis x=10,
            ymode=log,log basis y=10,
            height=.2\textheight,
            width=.97\textwidth,
            xmin=32,
            xmax=1073741824,
            ymax=1.5E-8,
            ymin=1E-10,
            axis lines*=left,
            legend style={at={(1.7,0.5)}, anchor=east,legend columns=1, font=\scriptsize},
            ylabel={Time in s/ n},
            xlabel={Number of values n},
            yticklabel style = {font=\scriptsize,xshift=0.5ex},
            xticklabel style = {font=\scriptsize,yshift=0.5ex},
            ylabel style = {font=\scriptsize},
            xlabel style = {font=\scriptsize},
            every node near coord/.append style={font=\scriptsize, color=gray,yshift=+1pt, /pgf/number format/.cd,fixed zerofill,precision=0},
                 every axis legend/.append style={nodes={right}},
        ]
\addplot[draw=purple] table [x index=0, y index=2, header=true] {./resultsres-intel1partitions-double.data}; 
\addplot[draw=ForestGreen] table [x index=0, y index=6, header=true] {./resultsres-intel1partitions-double.data}; 
\addplot[draw=ForestGreen, only marks, mark=o,nodes near coords,point meta=explicit symbolic] table [x index=0, y index=6, meta index=7, header=true] {./resultsres-intel1partitions-double.data};
    \end{axis}
    \end{tikzpicture}
        \caption{Floating-point (double)}
    \end{subfigure}
    \hfill
    \begin{subfigure}[c]{\columnwidth}
    \begin{tikzpicture}
    \begin{axis}[
            xmode=log,log basis x=10,
            ymode=log,log basis y=10,
            height=.2\textheight,
            width=.97\textwidth,
            xmin=32,
            xmax=1073741824,
            ymax=1.5E-8,
            ymin=1E-10,
            axis lines*=left,
            legend style={at={(1.7,0.5)}, anchor=east,legend columns=1, font=\scriptsize},
            ylabel={Time in s/ n},
            xlabel={Number of values n},
            yticklabel style = {font=\scriptsize,xshift=0.5ex},
            xticklabel style = {font=\scriptsize,yshift=0.5ex},
            ylabel style = {font=\scriptsize},
            xlabel style = {font=\scriptsize},
            every node near coord/.append style={font=\scriptsize, color=gray,yshift=+1pt, /pgf/number format/.cd,fixed zerofill,precision=0},
                 every axis legend/.append style={nodes={right}},
        ]
\addplot[draw=purple] table [x index=0, y index=2, header=true] {./resultsres-intel1partitions-pair-int.data}; 
\addplot[draw=ForestGreen] table [x index=0, y index=4, header=true] {./resultsres-intel1partitions-pair-int.data}; 
\addplot[draw=ForestGreen, only marks, mark=o,nodes near coords,point meta=explicit symbolic] table [x index=0, y index=4, meta index=5, header=true] {./resultsres-intel1partitions-pair-int.data};
    \end{axis}
    \end{tikzpicture}
    \caption{Key/value integer pair (int[2])}
    \end{subfigure}
    \caption{Execution time divided by \emph{n} of elements to partition arrays filled with random values with sizes from 2$^1$ to 2$^{30}$ ($\approx 10^9$).
            The pivot is selected randomly. the AVX-512-partition line.
            The execution time is obtained from the average of 20 executions.
            The speedup of the AVX-512-partition against the STL is shown above.}
\label{res:partitioning}
\end{figure}
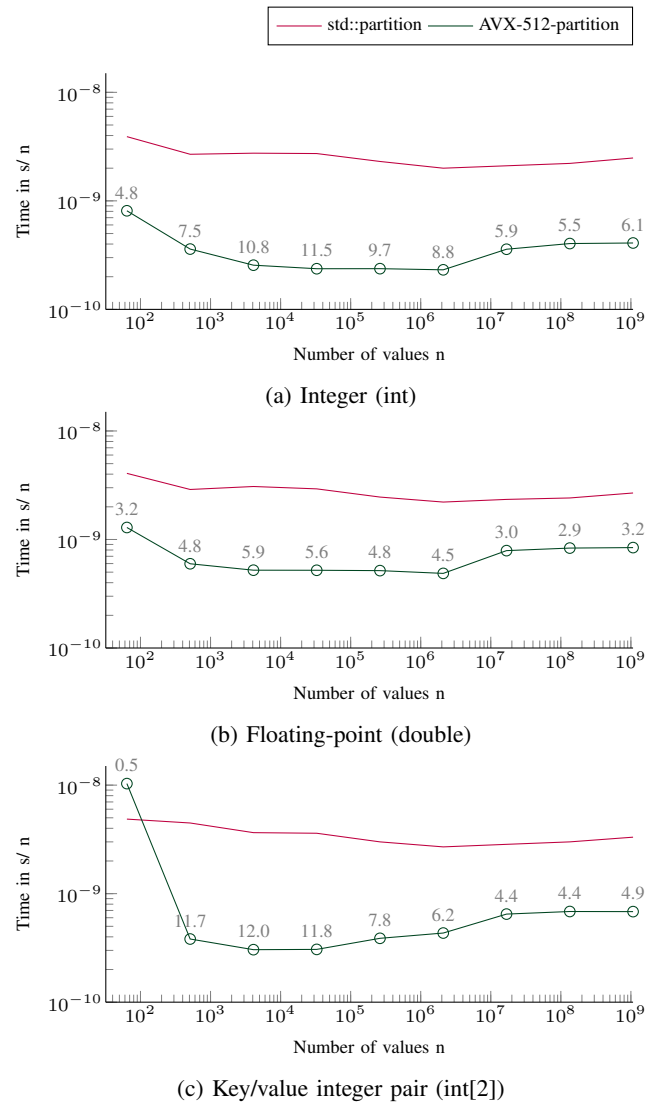

Figure~\ref{res:partitioning} shows the execution times to partition using our AVX-512-partition or the STL's partition function.
Our method provides again a speedup of an average factor of 4.
For the three configurations, an overhead impacts our implementation and the STL when partitioning arrays larger than 10$^7$ items.
Our AVX-512-partition remains faster, but its speedup decreases from 4 to 3.
This phenomena is related to cache effects since 10$^7$ integers values occupy 40M-Bytes, which is more than the L3 cache size.
In addition, we see that this effect starts from 10$^5$ when partitioning key/value pairs.

\subsection{Performance to Sort Large Arrays}

\begin{figure}
\begin{subfigure}[c]{\columnwidth}
        \begin{tikzpicture}
        \begin{axis}[
            xmode=log,log basis x=10,
            ymode=log,log basis y=10,
            height=.2\textheight,
            width=.97\textwidth,
            xmin=32,
            xmax=1073741824,
            ymax=1.5E-8,
            ymin=1.9E-10,
            axis lines*=left,
            legend style={at={(1,1.2)}, anchor=east,legend columns=3, font=\scriptsize},
            ylabel={Time in s/ n log(n)},
            xlabel={Number of values n},
            yticklabel style = {font=\scriptsize,xshift=0.5ex},
            xticklabel style = {font=\scriptsize,yshift=0.5ex},
            ylabel style = {font=\scriptsize},
            xlabel style = {font=\scriptsize},
            every node near coord/.append style={font=\scriptsize, color=gray,yshift=+1pt, /pgf/number format/.cd,fixed zerofill,precision=0},
                 every axis legend/.append style={nodes={right}},
            ]
\addplot[draw=purple] table [x index=0, y index=2, header=true] {./resultsres-intel1res-int.data}; 
\addplot[draw=brown] table [x index=0, y index=14, header=true] {./resultsres-intel1res-int.data}; 
\addplot[draw=ForestGreen] table [x index=0, y index=12, header=true] {./resultsres-intel1res-int.data}; 
\addplot[draw=ForestGreen, only marks, mark=o,nodes near coords,point meta=explicit symbolic] table [x index=0, y index=12, meta index=15, header=true] {./resultsres-intel1res-int.data};

\legend{std::sort, Intel IPP, AVX-512-QS}

        \end{axis}
        \end{tikzpicture}
    \caption{Integer (int)}
    \end{subfigure}
    \hfill
    \begin{subfigure}[c]{\columnwidth}
    \begin{tikzpicture}
    \begin{axis}[
            xmode=log,log basis x=10,
            ymode=log,log basis y=10,
            height=.2\textheight,
            width=.97\textwidth,
            xmin=32,
            xmax=1073741824,
            ymax=1.5E-8,
            ymin=1.9E-10,
            axis lines*=left,
            legend style={at={(1.7,0.5)}, anchor=east,legend columns=1, font=\scriptsize},
            ylabel={Time in s/ n log(n)},
            xlabel={Number of values n},
            yticklabel style = {font=\scriptsize,xshift=0.5ex},
            xticklabel style = {font=\scriptsize,yshift=0.5ex},
            ylabel style = {font=\scriptsize},
            xlabel style = {font=\scriptsize},
            every node near coord/.append style={font=\scriptsize, color=gray,yshift=+1pt, /pgf/number format/.cd,fixed zerofill,precision=0},
                 every axis legend/.append style={nodes={right}},
        ]
\addplot[draw=purple] table [x index=0, y index=2, header=true] {./resultsres-intel1res-double.data}; 
\addplot[draw=brown] table [x index=0, y index=14, header=true] {./resultsres-intel1res-double.data}; 
\addplot[draw=ForestGreen] table [x index=0, y index=12, header=true] {./resultsres-intel1res-double.data}; 
\addplot[draw=ForestGreen, only marks, mark=o,nodes near coords,point meta=explicit symbolic] table [x index=0, y index=12, meta index=15, header=true] {./resultsres-intel1res-double.data};
    \end{axis}
    \end{tikzpicture}
        \caption{Floating-point (double)}
    \end{subfigure}
    \hfill
    \begin{subfigure}[c]{\columnwidth}
    \begin{tikzpicture}
    \begin{axis}[
            xmode=log,log basis x=10,
            ymode=log,log basis y=10,
            height=.2\textheight,
            width=.97\textwidth,
            xmin=32,
            xmax=1073741824,
            ymax=1.5E-8,
            ymin=1.9E-10,
            axis lines*=left,
            legend style={at={(1.7,0.5)}, anchor=east,legend columns=1, font=\scriptsize},
            ylabel={Time in s/ n log(n)},
            xlabel={Number of values n},
            yticklabel style = {font=\scriptsize,xshift=0.5ex},
            xticklabel style = {font=\scriptsize,yshift=0.5ex},
            ylabel style = {font=\scriptsize},
            xlabel style = {font=\scriptsize},
            every node near coord/.append style={font=\scriptsize, color=gray,yshift=+1pt, /pgf/number format/.cd,fixed zerofill,precision=0},
                 every axis legend/.append style={nodes={right}},
        ]
\addplot[draw=purple] table [x index=0, y index=2, header=true] {./resultsres-intel1res-pair-int.data}; 
\addplot[draw=ForestGreen] table [x index=0, y index=4, header=true] {./resultsres-intel1res-pair-int.data}; 
\addplot[draw=ForestGreen, only marks, mark=o,nodes near coords,point meta=explicit symbolic] table [x index=0, y index=4, meta index=5, header=true] {./resultsres-intel1res-pair-int.data}; 
    \end{axis}
    \end{tikzpicture}
    \caption{Key/value integer pair (int[2])}
    \end{subfigure}
    \caption{Execution time divided by \emph{n log(n)} to sort arrays filled with random values with sizes from 2$^1$ to 2$^{30}$ ($\approx 10^9$).
            The execution time is obtained from the average of 5 executions.
            The speedup of the AVX-512-bitonic against the fastest between STL and IPP is shown above the AVX-512-bitonic line.}
\label{res:largesize}
\end{figure}
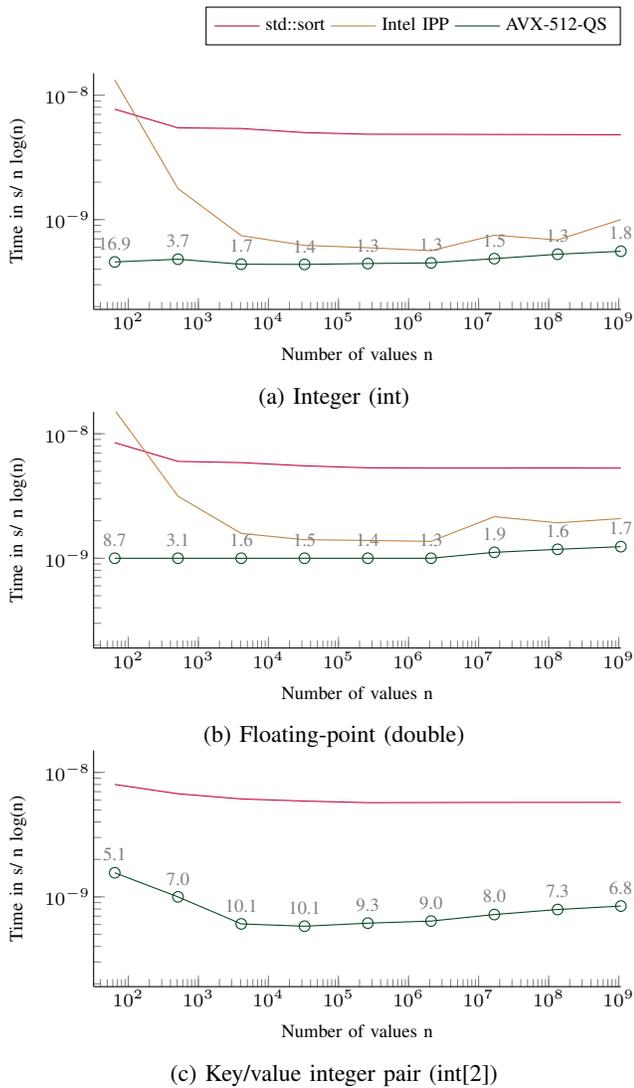

Figure~\ref{res:largesize} shows the execution times to sort arrays up to a size of 10$^9$ items.
Our AVX-512-QS is always faster in all configurations.
The difference between AVX-512-QS and the STL sort seems stable for any size with a speedup of more than 6 to our benefit.
However, while the Intel IPP is not efficient for arrays with less than 10$^4$ elements, its performance is really close to the AVX-512-QS for large arrays.
The same effect found when partitioning appears when sorting arrays larger than 10$^7$ items.
All three sorting functions are impacted, but the IPP seems more slowdown than our method, because it is based on a different access pattern, such that the AVX-512-QS is almost twice as fast as IPP for a size of 10$^9$ items.

\FloatBarrier
\section{Conclusions}
\label{sec:conclusion}

In this paper, we introduced new Bitonic sort and a new partition algorithm that have been designed for the AVX-512 instruction set.
These two functions are used in our Quicksort variant which makes it possible to have a fully vectorized implementation (at the exception of partitioning tiny arrays).
Our approach shows superior performance on Intel SKL in all configurations against two reference libraries: the GNU \emph{C++} STL, and the Intel IPP. 
It provides a speedup of 8 to sort small arrays (less than 16 SIMD-vectors), and a speedup of 4 and 1.4 for large arrays, against the \emph{C++} STL and the Intel IPP, respectively.
These results should also motivate the community to revisit common problems, because some algorithms may become competitive by being vectorizable, or improved, thanks to AVX-512's novelties.
Our source code is publicly available and ready to be used and compared.
In the future, we intend to design a parallel implementation of our AVX-512-QS, and we expect the recursive partitioning to be naturally parallelized with a task-based scheme on top of OpenMP.


\appendix
\section{APPENDIX}

\subsection{Scalar Quicksort Algorithm}
 \label{algo:qs}

\begin{algorithm2e}[h!]
\SetAlgoLined
\DontPrintSemicolon
\KwIn{array: an array to sort. length: the size of array.}
\KwOut{array: the array sorted.}
 \Fn{QS(array, length)}
 {
    QS\_core(array, 0, length-1)\;
 }
\BlankLine
 \Fn{QS\_core(array, left, right)}
 {
    \If{ left $<$ right }
    {
        \tcp{Naive method, select value in the middle}
        pivot\_idx = ((right-left)/2) + left\;
        swap(array[pivot\_idx], array[right])\;
        partition\_bound = partition(array, left, right, array[right])\;
        swap(array[partition\_bound], array[right])\;
        QS\_core(array, left, partition\_bound-1)\;
        QS\_core(array, partition\_bound+1, right)\;
    }
 }
\BlankLine
 \Fn{partition(array, left, right, pivot\_value)}
 {
    \For{idx\_read $\leftarrow$ left \KwTo right}
    {
        \If{array[idx\_read] < pivot\_value}
        {
            swap(array[idx\_read], array[left])\;
            left += 1\;
        }
    }
    return left;
 }
 \caption{Quicksort}
\end{algorithm2e}

\subsection{Source Code Extracts}
 \label{app:code}

\begin{minipage}{\linewidth}
\begin{lstlisting}[linewidth=\columnwidth,label={code:bitonic1v}, caption={AVX-512 Bitonic sort for one simd-vector of double floating-point values.}]
inline __m512d AVX_512_bitonic_sort_1v(__m512d input){
{
    __m512i idxNoNeigh = _mm512_set_epi64(6, 7, 4, 5, 2, 3, 0, 1);
    __m512d permNeigh = _mm512_permutexvar_pd(idxNoNeigh, input);
    __m512d permNeighMin = _mm512_min_pd(permNeigh, input);
    __m512d permNeighMax = _mm512_max_pd(permNeigh, input);
    input = _mm512_mask_mov_pd(permNeighMin, 0xAA, permNeighMax);
}
{
    __m512i idxNoNeigh = _mm512_set_epi64(4, 5, 6, 7, 0, 1, 2, 3);
    __m512d permNeigh = _mm512_permutexvar_pd(idxNoNeigh, input);
    __m512d permNeighMin = _mm512_min_pd(permNeigh, input);
    __m512d permNeighMax = _mm512_max_pd(permNeigh, input);
    input = _mm512_mask_mov_pd(permNeighMin, 0xCC, permNeighMax);
}
{
    __m512i idxNoNeigh = _mm512_set_epi64(6, 7, 4, 5, 2, 3, 0, 1);
    __m512d permNeigh = _mm512_permutexvar_pd(idxNoNeigh, input);
    __m512d permNeighMin = _mm512_min_pd(permNeigh, input);
    __m512d permNeighMax = _mm512_max_pd(permNeigh, input);
    input = _mm512_mask_mov_pd(permNeighMin, 0xAA, permNeighMax);
}
{
    __m512i idxNoNeigh = _mm512_set_epi64(0, 1, 2, 3, 4, 5, 6, 7);
    __m512d permNeigh = _mm512_permutexvar_pd(idxNoNeigh, input);
    __m512d permNeighMin = _mm512_min_pd(permNeigh, input);
    __m512d permNeighMax = _mm512_max_pd(permNeigh, input);
    input = _mm512_mask_mov_pd(permNeighMin, 0xF0, permNeighMax);
}
{
    __m512i idxNoNeigh = _mm512_set_epi64(5, 4, 7, 6, 1, 0, 3, 2);
    __m512d permNeigh = _mm512_permutexvar_pd(idxNoNeigh, input);
    __m512d permNeighMin = _mm512_min_pd(permNeigh, input);
    __m512d permNeighMax = _mm512_max_pd(permNeigh, input);
    input = _mm512_mask_mov_pd(permNeighMin, 0xCC, permNeighMax);
}
{
    __m512i idxNoNeigh = _mm512_set_epi64(6, 7, 4, 5, 2, 3, 0, 1);
    __m512d permNeigh = _mm512_permutexvar_pd(idxNoNeigh, input);
    __m512d permNeighMin = _mm512_min_pd(permNeigh, input);
    __m512d permNeighMax = _mm512_max_pd(permNeigh, input);
    input = _mm512_mask_mov_pd(permNeighMin, 0xAA, permNeighMax);
}

return input;
}
  \end{lstlisting}
\end{minipage}

\begin{minipage}{\linewidth}
\begin{lstlisting}[linewidth=\columnwidth,label={code:simdpartition512}, caption={AVX-512 partitioning of a double floating-point array (AVX-512-partition)}]
template <class IndexType>
static inline IndexType AVX_512_partition(double array[], IndexType left, IndexType right, const double pivot){
const IndexType S = 8;//(512/8)/sizeof(double);

if(right-left+1 < 2*S){
    return CoreScalarPartition<double,IndexType>(array, left, right, pivot);
}

__m512d pivotvec = _mm512_set1_pd(pivot);

__m512d left_val = _mm512_loadu_pd(&array[left]);
IndexType left_w = left;
left += S;

IndexType right_w = right+1;
right -= S-1;
__m512d right_val = _mm512_loadu_pd(&array[right]);

while(left + S <= right){
    const IndexType free_left = left - left_w;
    const IndexType free_right = right_w - right;

    __m512d val;
    if( free_left <= free_right ){
        val = _mm512_loadu_pd(&array[left]);
        left += S;
    }
    else{
        right -= S;
        val = _mm512_loadu_pd(&array[right]);
    }

    __mmask8 mask = _mm512_cmp_pd_mask(val, pivotvec, _CMP_LE_OQ);

    const IndexType nb_low = popcount(mask);
    const IndexType nb_high = S-nb_low;

    _mm512_mask_compressstoreu_pd(&array[left_w],mask,val);
    left_w += nb_low;

    right_w -= nb_high;
    _mm512_mask_compressstoreu_pd(&array[right_w],~mask,val);
}

{
    const IndexType remaining = right - left;
    __m512d val = _mm512_loadu_pd(&array[left]);
    left = right;

    __mmask8 mask = _mm512_cmp_pd_mask(val, pivotvec, _CMP_LE_OQ);

    __mmask8 mask_low = mask & ~(0xFF << remaining);
    __mmask8 mask_high = (~mask) & ~(0xFF << remaining);

    const IndexType nb_low = popcount(mask_low);
    const IndexType nb_high = popcount(mask_high);

    _mm512_mask_compressstoreu_pd(&array[left_w],mask_low,val);
    left_w += nb_low;

    right_w -= nb_high;
    _mm512_mask_compressstoreu_pd(&array[right_w],mask_high,val);
}
{
    __mmask8 mask = _mm512_cmp_pd_mask(left_val, pivotvec, _CMP_LE_OQ);

    const IndexType nb_low = popcount(mask);
    const IndexType nb_high = S-nb_low;

    _mm512_mask_compressstoreu_pd(&array[left_w],mask,left_val);
    left_w += nb_low;

    right_w -= nb_high;
    _mm512_mask_compressstoreu_pd(&array[right_w],~mask,left_val);
}
{
    __mmask8 mask = _mm512_cmp_pd_mask(right_val, pivotvec, _CMP_LE_OQ);

    const IndexType nb_low = popcount(mask);
    const IndexType nb_high = S-nb_low;

    _mm512_mask_compressstoreu_pd(&array[left_w],mask,right_val);
    left_w += nb_low;

    right_w -= nb_high;
    _mm512_mask_compressstoreu_pd(&array[right_w],~mask,right_val);
}
return left_w;
}
  \end{lstlisting}
\end{minipage}

\FloatBarrier


\end{document}